# Title: Chiral Valley Edge States

**Authors:** Jian-Wei Liu[1,2,†], Gui-Geng Liu[3,4,†], Bo Zhang[1,†], Hao-Chang Mo[1], Ruifeng Li[2], Mingwei Li[1], Xiao-Dong Chen[1], Baile Zhang[2,5,*], Wen-Jie Chen[1,*], and Jian-Wen Dong[1,*]

**Affiliations:**
[1]School of Physics & State Key Laboratory of Optoelectronic Materials and Technologies, Sun Yat-sen University, Guangzhou 510275, China.
[2]Division of Physics and Applied Physics, School of Physical and Mathematical Sciences, Nanyang Technological University; Singapore, 637371, Singapore.
[3]Research Center for Industries of the Future, Westlake University; Hangzhou, 310030, China.
[4]Department of Electronic and Information Engineering, School of Engineering, Westlake University; Hangzhou, 310030, China.
[5]Centre for Disruptive Photonic Technologies, The Photonics Institute, Nanyang Technological University; Singapore, 637371, Singapore.
†These authors contributed equally to this work.
*Corresponding authors. Email: blzhang@ntu.edu.sg;
chenwenj5@mail.sysu.edu.cn;
dongjwen@mail.sysu.edu.cn.

**Abstract:**
  Valleytronics has emerged as a promising paradigm, enabling comprehensive control of the valley degree of freedom (DoF) for energy-efficient and high-speed information processing. However, backscattering-induced valley depolarization remains a fundamental limitation, stemming from the weak topological protection of the valley Hall phase. Here, we propose and demonstrate the concept of chiral valley edge states, which integrate the robust unidirectional chiral edge states with valley DoF. By controlling the valley Dirac masses, we selectively confine the chiral edge band around a single valley, enabling back-scattering-free propagation while imparting valley polarization. Our strategy not only addresses the valley depolarization issue but also introduces a unique functionality—valley multiplexing—allowing independent and arbitrary control over waves associated with different valley polarizations. We demonstrate our concept experimentally within hybrid topological photonic crystal systems composed of Chern and valley photonic crystals. Moreover, two key components for valley multiplexing are demonstrated: a valley (de-)multiplexer and a valley-locked waveguide crossing, facilitating non-interfering signal routing. Our results establish a novel interplay between the topological quantum Hall and valley Hall phases, offering a new framework for robust valley-based information processing.

**Introduction**

Owing to advantages in energy efficiency, robustness, and high processing speed, valley has been highlighted as a promising degree of freedom (DoF) for information encoding and storage, catalyzing the birth of valleytronics [1-3] and valley photonics [4,5]. In certain crystalline materials like graphene and transition metal dichalcogenides (TMDCs), valleys refer to local extrema in energy bands, separated in momentum space [6]. To process information via valley DoF, it is crucial to enable valley wave transport and control valley polarization [7-10]. However, since information is encoded in momentum space, the inevitable defect-induced back-scattering is a persistent obstacle. In valleytronics, back-scattering leads to inter-valley scattering even with perfect valley initialization [11], causing severe valley depolarization and limiting valley lifetime; while in valley photonics, it gives rise to Anderson localization and propagation loss [12].

The limited robustness of valley polarization mainly originates from its association with a relatively weak topological phase—the valley Hall phase—which is protected solely by crystalline symmetry. Conversely, the quantum (anomalous) Hall phase has long been recognized as a strong topological phase, supporting unidirectional chiral edge states in accordance with the bulk-boundary correspondence [13-15]. Due to the broken time-reversal symmetry and the resulting non-reciprocity, these chiral edge states are entirely immune to back-scattering and remain unaffected by defects and interface deformations of any kind. This novel property of dissipation-less transmission offers the potential for addressing the power consumption issue inherent in traditional complementary metal-oxide-semiconductor (CMOS) device technologies, spurring research efforts into the Chern networks [16-18]. Nonetheless, while the existence of chiral edge states is guaranteed by topology, their dispersion configuration in momentum space lacks specific constraints. Consequently, these chiral edge states typically span the Brillouin zone (BZ), without exhibiting valley-dependent characteristics.

Here, we integrate chiral edge states with the valley DoF, proposing the concept of chiral valley edge states. In crystalline lattices, valleys can be effectively considered as low-energy massive Dirac quasiparticles, with their topological properties determined by the Dirac masses [19,20]. Thus, by means of Dirac mass engineering, we selectively confine the chiral edge band around a single valley, thereby imparting valley-dependent characteristics. Due to the back-scattering-free propagation, the chiral valley edge states can robustly preserve valley polarization during transmission, potentially overcoming the longstanding valley depolarization issue in valleytronics. Furthermore, via precisely engineering the spatial distribution of valley Dirac masses, we achieve independent and arbitrary control over waves with both valley polarizations, referred to as the valley multiplexing. To the best of our knowledge, this is a unique feature which is absent among previous valleytronic schemes, let alone in the Chern networks.

From a broader perspective, our proposed chiral valley edge states arise from the interplay between the quantum Hall phase and the valley Hall phase. To demonstrate this concept, we realize chiral valley edge states in hybrid topological photonic crystal systems composed of Chern photonic crystals (CPCs) and valley photonic crystals (VPCs). The robust valley polarization and unidirectional propagation are verified both numerically and exponentially. Moreover, to demonstrate the valley multiplexing, two representative valley-dependent photonic devices are exhibited: a photonic valley (de-)multiplexer, serving as the core component enabling valley multiplexing; and a photonic valley-locked waveguide crossing, which facilitates non-interfering crossing between waves with different valley-polarizations—a functionality that is difficult to achieve using either pure valleytronic or pure Chern systems.

**Result**
**Chiral valley edge states**

It is well established that topological quantum Hall insulators host unidirectional gapless chiral edge states, in accordance with the bulk-boundary correspondence (Fig. 1a) [21-23]. Nonetheless, these chiral edge states typically span both valleys in momentum space and thus lack valley-dependent characteristics. On the other hand, owing to the time-reversal symmetry, valley Hall insulators support "helical" valley edge states[24-26], which manifest as two counter-propagating edge modes with distinct valley polarizations (Fig. 1b). Their inherent reciprocity makes the back-scattering unavoidable.

Our primary objective in realizing the chiral valley edge state is to confine the chiral edge band around a single valley (Fig. 1c). In the following, we present an approach based on valley Dirac mass engineering, showing that chiral edge bands can be selectively restricted to a specific valley. These states integrate unidirectional propagation with valley-dependent characteristics, inheriting the advantages of both quantum Hall and valley Hall phases.

Usually in two-dimensional (2D) graphene-like lattices, valleys refer to a set of local extrema in energy bands, located at BZ corners (i.e., the K and K' points). The eigenstates near a given valley (say, the K valley) are governed by a local Dirac Hamiltonian $H_K(\bm{k}) = v_x k_x \sigma_x + v_y k_y \sigma_y + m_K \sigma_z$. As such, a valley can be effectively deemed as a massive Dirac quasiparticle, with its topological features characterized by its Dirac mass—$m_K$. At a domain wall between two lattices with opposite-sign $m_K$ (see Fig. 1d), a unidirectional K-valley Jackiw-Rebbi mode is topologically ensured, whose propagation direction is determined by the signs of two $m_K$ [27].

For a single 2D lattice, the Dirac masses on different valleys ($m_K$ and $m_{K'}$) can be separately engineered. This allows us to mold the flows of different valley states independently by spatially encoding the distributions of $m_K(\bm{r})$ and $m_{K'}(\bm{r})$, thereby enabling precise control over the valley index of chiral edge states. For instance, as shown in Fig. 1e, by engineering $m_K(\bm{r})$ but keeping a homogeneous $m_{K'}(\bm{r})$, we can construct a Z-shaped bending waveguide supporting chiral K-valley waves. Conversely, by keeping $m_K(\bm{r})$ uniform and engineering $m_{K'}(\bm{r})$, a back-scattering-immune power splitter for K'-valley waves can be created (Fig. 1f). More significantly, since the valley indices of each chiral edge mode can be independently controlled, it is possible to simultaneously implement a unidirectional K-polarized bending waveguide and a back-scattering-immune K'-polarized power splitter within the same structure by superposing K- and K'-valley masses (Fig. 1g). Moreover, via engineering the spatial distribution of Dirac masses, the flows of these two chiral valley edge states can be arbitrarily routed. It is worth emphasizing that, this flexible and comprehensive utilization of both valleys—referred to as valley multiplexing—not only multiplies the transmission capacity but also enables diverse functionalities on a single chip, offering significant potential for high-density integration.

Our proposed chiral valley edge states arise from the interplay between the quantum Hall phase and the valley Hall phase. Therefore, we realize the chiral valley edge states in hybrid topological photonic crystal systems composed of topological CPCs and VPCs, which together provide sufficient DoFs for Dirac mass engineering. For instance, Fig. 2a plots the band structure of CPC-1, composed of a honeycomb array of YIG rods. Figure 2b presents the band structure of VPC-1, made from two sets of dielectric rods with different radii. Berry curvatures of the two crystals are encoded in color, reflecting their different valley Dirac masses. Because their K (K') valley masses take opposite (same) signs, when we combine these two crystals together (Fig. 2c), the domain-wall between them supports only a chiral K-valley mode propagating to the right, as demonstrated by the simulated and measured edge dispersions in Fig. 2d. Similarly, we can selectively endow the chiral edge state with K'-valley polarization by utilizing CPC-2 (the time-reversal counterpart of CPC-1, featuring positive masses at both valleys) and VPC-1 [see Supplementary Figs. S2 and S3 for detailed edge dispersion].

We notice that such type of chiral valley edge states has recently been proposed in condensed matter and photonic systems, where they termed as the perfect valley filters [28,29]. However, in these works, they only theoretically exhibited a limited part of using these states as a filter. The more promising potential of controlling the valley polarization by valley Dirac masses has yet to be mentioned, not to mention the unique valley multiplexing feature.

To further present the unique valley multiplexing functionality, two representative photonic devices based on the chiral valley edge states are demonstrated: a photonic valley (de-)multiplexer, serving as the core component enabling valley multiplexing; and a photonic valley-locked waveguide crossing, which facilitates non-interfering crossing between waves with different valley-polarizations.

**Photonic valley (de-)multiplexer**

Multiplexer, as a core component of multiplexing techniques [30,31], is designed to combine individual signals into a composite one. Transmitting such a composite signal through a shared bus channel not only enhances data transmission capacity but also improves chip space utilization. In the context of valley multiplexing, which aims to merge K- and K'-polarized waves into a single waveguide (i.e., the bus channel), the Dirac mass distributions for two valleys can be deduced as in Fig. 2e. Consequently, based on the superposition scheme, a photonic valley multiplexer can then be realized, as illustrated in Fig. 2f, forming a Y-junction between three domain-wall waveguides. We designed two types of CPCs and one VPC accordingly, tailored to achieve the desired Dirac mass configurations.

Since CPC-1 and CPC-2 have opposite masses at both valleys, their domain-wall supports both K and K'-valley propagating modes, serving as a shared bus waveguide [see Supplementary Fig. S4 for edge dispersion]. Then we can construct the photonic valley multiplexer. As simulated in Figs. 2g and 2h, the downward-propagating K-polarized wave and upward-propagating K'-polarized wave converge into the horizontal bus waveguide. More importantly, these two valley modes propagate independently without interference due to their different valley polarization (insets of Figs. 2f and 2g). Field profiles in Fig. 2f and 2g confirm that waves from the upper and lower channels can merge into the bus waveguide without any valley flipping, verifying not only the functionality of valley multiplexer but also the one-way characteristic of valley chiral edge states.

For independent manipulation of valley-polarized modes, a valley de-multiplexer is essential to split different valley waves into designated channels. Notably, since all three channels in Fig. 2f are one-way waveguides, this structure cannot be used as a de-multiplexer by simply impinging waves from the right-hand side. Instead, a photonic valley de-multiplexer can be implemented by its reversed structure as depicted in Fig. 3a. When electromagnetic waves are incident from the left, the two valley components are routed into different channels at the Y-junction (Fig. 3b), demonstrating de-multiplexing functionality. To verify this valley-dependent splitting effect, we compare the $E_z$-field profiles with the simulated eigen edge modes. As shown in Figs. 3c and 3d, the upper waveguide field is asymmetric about the edge with a clockwise energy vortex, while the field in lower waveguide is symmetric with two energy vortexes. These features align well with the eigen mode at K/K' valley, confirming their valley polarization purity.

The two channels' pure valley polarizations are also manifested in their topological refractions at the exits. As predicted by $k$-space analysis in Fig. 3e, the measured output beam at the upper-right exit is refracted to the left while the beam at the lower-right exit is refracted to the right. Two acrylic semi-circular lenses are placed at the exits to couple the beams outward. Moreover, the propagation paths of two valley-polarized waves can be well controlled by rearranging the Dirac mass distributions. As simulated and measured in Fig. 3f, by reversing the valley Dirac masses of VPC domain (denoted as VPC-2), the propagation paths of the K- and K'-polarized modes are flipped, enabling flexible routing.

**Valley-locked waveguide crossing**

Apart from multiplexer and de-multiplexer, waveguide crossing is another critical component for high-density integrated in photonic chips. As device density and circuit complexity increase, optical waveguide crossings become essential elements that significantly influence the overall performance of photonic chips [32-36]. Here, we propose a valley-locked waveguide crossing based on valley-dependent Dirac mass engineering. To our best knowledge, crossing between topological waveguides has yet to be proposed or realized so far, whose performance can be effectively improved by topological aspects of robust edge states.

In order to have a K-valley edge mode propagating horizontally and a K'-valley edge mode propagating vertically, we design the Dirac mass distributions for two valleys separately (Fig. 4a), and then construct the valley-locked waveguide crossing via the superposition scheme. As a result, the valley Dirac mass configuration required for each domain is illustrated in Fig. 4b: CPC-1 (blue) and CPC-2 (pink) occupy the upper-left and lower-right regions, while VPC-1 (yellow) and VPC-2 (orange) occupy the lower-left and upper-right. Due to the distinct valley polarizations of edge modes, the two channels are decoupled from each other even at the intersection point, resulting in minimal crosstalk. Furthermore, inheriting the topological protection of chiral edge states, this type of waveguide crossing can effectively minimize both back-scattering loss and crosstalk.

We validate the performance of the valley-locked waveguide crossing by analyzing its transmission spectra and electromagnetic field profiles. First, we launch electromagnetic waves from port 1, then probe the transmission spectra from all other ports, as shown in Fig. 4c. The red line plots the crosstalk from port 1 to port 4, which is lower than -16.2 dB within the edge-mode-supported frequency range (highlighted in gray, from 5.20 to 5.67 GHz). Notably, the green line shows even lower crosstalk from port 1 to port 3, confirming the unidirectional nature of our waveguide channels. Corresponding experimental measurements of transmission spectra are shown in Fig. 4d, indicating an average crosstalk level of approximately -9.8 dB. We scan the near-field pattern to observe the cross transmission of two valley waves. The measured field profile at 5.57 GHz demonstrates a smooth propagation with negligible crosstalk (Fig. 4e). Besides, a 2D Fourier transform is applied to the measured field profiles to obtain its momentum information. As demonstrated in Fig. 4f, the momentum is concentrated around the K valley, confirming the valley-locked feature of topological waveguide crossing. Similar behaviors are observed when electromagnetic waves are excited from port 3, showing a crosstalk level below -15.7 dB in simulation (Fig. 4g) and an average crosstalk about -11.8 dB in experiment (Fig. 4h). Its momentum distribution concentrates around K' valley (Fig. 4j). These findings demonstrate that signals in the two channels maintain distinct valley polarizations, effectively ensuring the low crosstalk in this topological waveguide crossing.

**Discussion**

In this work, we propose the concept of chiral valley edge states, which integrate chiral edge states with the valley degree of freedom. Our approach is based on Dirac mass engineering for individual valleys, enabling selective confinement of the chiral edge band around a single valley and thus imparting valley-dependent properties. Via precisely configuring the spatial distribution of Dirac mass for each valley, we achieve independent control of both valley states along arbitrary pathways, thereby realizing valley multiplexing. To demonstrate this mechanism, we implement two key valley-dependent devices using topological photonic crystals. One is a photonic valley (de-)multiplexer, which is a crucial component for separating or converging modes with distinct valley polarizations. The other is a photonic valley-locked waveguide crossing, which can significantly enhance the performance of densely integrated photonic chips. Both devices are experimentally characterized in the microwave regime.

Although our concept is validated on a photonic crystal platform, the underlying design principle is general and can be readily extended to other physical systems, including condensed matter, acoustics and electronic circuits. For example, in condensed matter systems, chiral valley edge states could be realized by combining valleytronic and anomalous Hall materials[37-40]. On the other hand, in photonics, while realizing time-reversal breaking in visible regime may pose challenges due to the weak magneto-optic effects of gyrotropic materials, our scheme could be adapted to the terahertz regime [41-45]. We believe that our approach not only addresses the critical limitations in existing valleytronics and Chern networks but also offers promising opportunities for integrating valleytronics concepts into next-generation communication systems. Ultimately, it provides a powerful pathway for substantially boosting transmission capacity through comprehensive utilization of the valley degree of freedom.

**Methods**

**Materials.** The valley photonic crystals in this work are realized by two kinds of dielectric rods with radii $d_1 = 6 \text{ mm}$ and $d_2 = 7.5 \text{ mm}$, arranged in hexagonal lattice with lattice constant $a = 18 \text{ mm}$. The material of the rod is ceramic, with relative permittivity of 8.8 and relative permeability of 1. The Chern photonic crystals in this work are realized by using gyromagnetic rods biased by an external magnetic field. The gyromagnetic rods have radii of $d_0 = 6 \text{ mm}$, and also are arranged in hexagonal lattice with lattice constant $a = 18 \text{ m}$. The gyromagnetic material is yttrium iron garnet (YIG), a ferrite with relative permittivity 14.6. The relative magnetic permeability of the YIG has the form of

$$\boldsymbol{\mu} = \begin{bmatrix} \mu_r & i\mu_k & 0 \\ -i\mu_k & \mu_r & 0 \\ 0 & 0 & 1 \end{bmatrix}, \quad (1)$$

where $\mu_r = 1 + (\omega_0 + i\alpha\omega)\omega_m/((\omega_0 + i\alpha\omega)^2 - \omega^2)$, $\mu_k = \omega\omega_m/((\omega_0 + i\alpha\omega)^2 - \omega^2)$, $\omega_m = \gamma M_s$, $\omega_0 = \gamma H_0$, $M_s = 0.195 \text{ T}$ is the saturation magnetization, $H_0 = 0.042\text{T}$ is the external magnetic field, $\gamma = 1.76 \times 10^{11} \text{ s}^{-1}\text{T}^{-1}$ is the gyromagnetic ratio, $\alpha = 0.0001$ is the damping coefficient, and $\omega$ is the operating frequency. The Dirac masses of the Chern photonic crystals can be controlled by the direction of the external magnetic field, which can be tuned by the direction of magnets in the experiment. Both the ceramic and YIG rods have a height of 5 mm, and are sandwiched between two metallic plates to mimic the two-dimensional condition.

**Simulation.** The dispersion relations and field patterns are simulated using the finite element software COMSOL Multiphysics. The unit cell of the Chern photonic crystals and valley photonic crystals are hexagonal lattice, of which we applied Floquet boundary conditions on the outer boundaries when calculating the bulk band structures in Fig. 2a and 2b. The edge band structure in Fig. 2d is calculated using a supercell with $1 \times 14$ periods.

**References**
[1] S. A. Vitale, D. Nezich, J. O. Varghese, P. Kim, N. Gedik, P. Jarillo-Herrero, D. Xiao, and M. Rothschild, Valleytronics: Opportunities, Challenges, and Paths Forward, Small **14**, e1801483 (2018).
[2] A. Rasmita and W.-b. Gao, Opto-valleytronics in the 2D van der Waals heterostructure, Nano Research **14**, 1901 (2020).
[3] A. Ciarrocchi, F. Tagarelli, A. Avsar, and A. Kis, Excitonic devices with van der Waals heterostructures: valleytronics meets twistronics, Nature Reviews Materials **7**, 449 (2022).
[4] H. Xue, Y. Yang, and B. Zhang, Topological Valley Photonics: Physics and Device Applications, Advanced Photonics Research **2**, 2100013 (2021).
[5] J.-W. Liu, F.-L. Shi, X.-T. He, G.-J. Tang, W.-J. Chen, X.-D. Chen, and J.-W. Dong, Valley photonic crystals, Advances in Physics: X **6**, 1905546 (2021).
[6] J. R. Schaibley, H. Yu, G. Clark, P. Rivera, J. S. Ross, K. L. Seyler, W. Yao, and X. Xu, Valleytronics in 2D materials, Nature Reviews Materials **1**, 16055 (2016).
[7] N. Rohling and G. Burkard, Universal quantum computing with spin and valley states, New Journal of Physics **14**, 083008 (2012).
[8] D. Culcer, A. L. Saraiva, B. Koiller, X. Hu, and S. Das Sarma, Valley-based noise-resistant quantum computation using Si quantum dots, Physical Review Letters **108**, 126804 (2012).
[9] Á. Jiménez-Galán, R. E. F. Silva, O. Smirnova, and M. Ivanov, Lightwave control of topological properties in 2D materials for sub-cycle and non-resonant valley manipulation, Nature Photonics **14**, 728 (2020).
[10] I. Tyulnev *et al.*, Valleytronics in bulk MoS(2) with a topologic optical field, Nature **628**, 746 (2024).
[11] R. Xu, Z. Zhang, J. Liang, and H. Zhu, Valleytronics: Fundamental Challenges and Materials Beyond Transition Metal Chalcogenides, Small, e2402139 (2024).
[12] C. A. Rosiek, G. Arregui, A. Vladimirova, M. Albrechtsen, B. Vosoughi Lahijani, R. E. Christiansen, and S. Stobbe, Observation of strong backscattering in valley-Hall photonic topological interface modes, Nature Photonics **17**, 386 (2023).
[13] F. D. Haldane, Model for a quantum Hall effect without Landau levels: Condensed-matter realization of the "parity anomaly", Physical Review Letters **61**, 2015 (1988).

[14] Z. Wang, Y. Chong, J. D. Joannopoulos, and M. Soljacic, Observation of unidirectional backscattering-immune topological electromagnetic states, Nature **461**, 772 (2009).

[15] C.-Z. Chang, C.-X. Liu, and A. H. MacDonald, Colloquium: Quantum anomalous Hall effect, Reviews of Modern Physics **95**, 011002 (2023).

[16] D. Ovchinnikov *et al.*, Topological current divider in a Chern insulator junction, Nature Communications **13**, 5967 (2022).

[17] Y. F. Zhao, R. Zhang, J. Cai, D. Zhuo, L. J. Zhou, Z. J. Yan, M. H. W. Chan, X. Xu, and C. Z. Chang, Creation of chiral interface channels for quantized transport in magnetic topological insulator multilayer heterostructures, Nature Communications **14**, 770 (2023).

[18] W. Yuan *et al.*, Electrical switching of the edge current chirality in quantum anomalous Hall insulators, Nature Materials **23**, 58 (2024).

[19] A. H. Castro Neto, F. Guinea, N. M. R. Peres, K. S. Novoselov, and A. K. Geim, The electronic properties of graphene, Reviews of Modern Physics **81**, 109 (2009).

[20] A. Bansil, H. Lin, and T. Das, Colloquium: Topological band theory, Reviews of Modern Physics **88**, 021004 (2016).

[21] Z. Wang, Y. Chong, J. Joannopoulos, and M. Soljačić, Reflection-free one-way edge modes in a gyromagnetic photonic crystal, Physical Review Letters **100**, 013905 (2008).

[22] Y. Poo, R.-x. Wu, Z. Lin, Y. Yang, and C. T. Chan, Experimental realization of self-guiding unidirectional electromagnetic edge states, Physical Review Letters **106**, 093903 (2011).

[23] B. B. A. Ndao, F. Vallini, A. E. Amili, Y. Fainman, and B. Kanté, Nonreciprocal lasing in topological cavities of arbitrary geometries, Science **358**, 636 (2017).

[24] T. Ma and G. Shvets, All-si valley-Hall photonic topological insulator, New Journal of Physics **18**, 025012 (2016).

[25] X.-D. Chen, F.-L. Zhao, M. Chen, and J.-W. Dong, Valley-contrasting physics in all-dielectric photonic crystals: Orbital angular momentum and topological propagation, Physical Review B **96**, 020202(R) (2017).

[26] F. Gao, H. Xue, Z. Yang, K. Lai, Y. Yu, X. Lin, Y. Chong, G. Shvets, and B. Zhang, Topologically protected refraction of robust kink states in valley photonic crystals, Nature Physics **14**, 140 (2017).

[27] R. Jackiw and C. Rebbi, Solitons with fermion number ½, Physical Review D **13**, 3398 (1976).

[28] H. Pan, X. Li, F. Zhang, and S. A. Yang, Perfect valley filter in a topological domain wall, Physical Review B **92**, 041404(R) (2015).

[29] X. Zhang, S. Li, Z. Lan, W. Gao, and M. L. N. Chen, Reconfigurable Photonic Valley Filter in Hybrid Topological Heterostructures, Laser & Photonics Reviews (2024).

[30] D. Dai and J. E. Bowers, Silicon-based on-chip multiplexing technologies and devices for Peta-bit optical interconnects, Nanophotonics **3**, 283 (2014).

[31] S. N. Khonina, N. L. Kazanskiy, M. A. Butt, and S. V. Karpeev, Optical multiplexing techniques and their marriage for on-chip and optical fiber communication: a review, Opto-Electronic Advances **5**, 210127 (2022).
[32] S. Wu, X. Mu, L. Cheng, S. Mao, and H. Y. Fu, State-of-the-Art and Perspectives on Silicon Waveguide Crossings: A Review, Micromachines (Basel) **11**, 326 (2020).
[33] Y. Ma, Y. Zhang, S. Yang, A. Novack, R. Ding, A. E. Lim, G. Q. Lo, T. Baehr-Jones, and M. Hochberg, Ultralow loss single layer submicron silicon waveguide crossing for SOI optical interconnect, Optics Express **21**, 29374 (2013).
[34] C. Sun, Y. Yu, and X. Zhang, Ultra-compact waveguide crossing for a mode-division multiplexing optical network, Optics Letters **42**, 4913 (2017).
[35] S. Li et al., Universal multimode waveguide crossing based on transformation optics, Optica **5**, 1549 (2018).
[36] H. Xu and Y. Shi, Metamaterial-Based Maxwell's Fisheye Lens for Multimode Waveguide Crossing, Laser & Photonics Reviews **12** (2018).
[37] K. F. Mak, K. L. McGill, J. Park, and P. L. McEuen, The valley Hall effect in MoS2 transistors, Science **344**, 1489 (2014).
[38] A. Vargas, F. Liu, C. Lane, D. Rubin, I. Bilgin, Z. Hennighausen, M. DeCapua, A. Bansil, and S. Kar, Tunable and laser-reconfigurable 2D heterocrystals obtained by epitaxial stacking of crystallographically incommensurate Bi2Se3 and MoS2 atomic layers, Science Advances **3**, e1601741 (2017).
[39] L. K. Rodenbach, I. T. Rosen, E. J. Fox, P. Zhang, L. Pan, K. L. Wang, M. A. Kastner, and D. Goldhaber-Gordon, Bulk dissipation in the quantum anomalous Hall effect, APL Materials **9**, 081116 (2021).
[40] I. T. Rosen, M. P. Andersen, L. K. Rodenbach, L. Tai, P. Zhang, K. L. Wang, M. A. Kastner, and D. Goldhaber-Gordon, Measured Potential Profile in a Quantum Anomalous Hall System Suggests Bulk-Dominated Current Flow, Physical Review Letters **129**, 246602 (2022).
[41] A. M. Shuvaev, G. V. Astakhov, A. Pimenov, C. Brune, H. Buhmann, and L. W. Molenkamp, Giant magneto-optical faraday effect in HgTe thin films in the terahertz spectral range, Physical Review Letters **106**, 107404 (2011).
[42] J. Chochol, K. Postava, M. Čada, M. Vanwolleghem, L. Halagačka, J.-F. Lampin, and J. Pištora, Magneto-optical properties of InSb for terahertz applications, AIP Advances **6**, 115021 (2016).
[43] D. Zhao, F. Fan, Z. Tan, H. Wang, and S. Chang, Tunable On-Chip Terahertz Isolator Based on Nonreciprocal Transverse Edge Spin State of Asymmetric Magneto-Plasmonic Waveguide, Laser & Photonics Reviews **17**, 2200509 (2022).
[44] D. Zhao, F. Fan, J. Liu, Z. Tan, H. Wang, Q. Yang, Q. Wen, and S. Chang, Terahertz magneto-optical metadevice for active spin-selective beam steering and energy distribution with nonreciprocal isolation, Optica **10**, 1295 (2023).
[45] R. Jia, T. C. Tan, S. S. Mishra, W. Wang, Y. J. Tan, and R. Singh, On-Chip Active Non-Reciprocal Topological Photonics, Advanced Materials, 2501711 (2025).

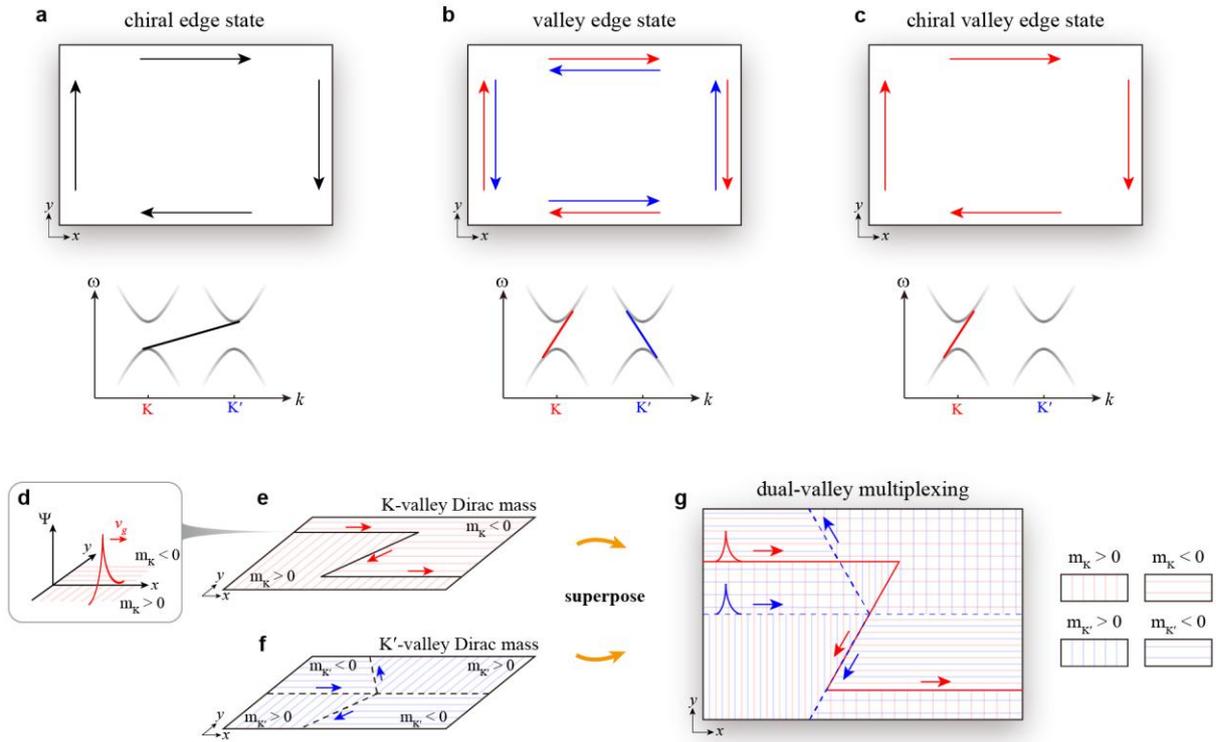

**Fig. 1 Chiral valley edge states. a** Chiral edge states. The edge band span both valleys and possesses no valley-dependent characteristics. **b** "helical" valley edge states with intrinsic reciprocity. The two counter-propagating edge waves contain distinct valley polarizations. **c** Chiral valley edge states, inheriting both one-way propagation property and valley-dependent characteristics. **d** Jackiw-Rebbi mode at the domain wall between two lattices with opposite Dirac masses. This mode propagates unidirectionally with its group velocity determined by the signs of Dirac masses on both sides. **e** Z-shaped bending waveguide for K-valley waves, realized by engineering the spatial distribution of $m_K(\mathbf{r})$. **f** A power splitter for K'-valley waves, designed via engineering the spatial distribution of $m_{K'}(\mathbf{r})$. Background strips represent valley Dirac masses in different domains: color denotes valley polarization, and strip orientation indicates the sign. **g** Dual-valley multiplexing achieved by precisely coding Dirac mass distributions at both valleys. The flow of both valley-polarized waves can be independently and flexibly molded.

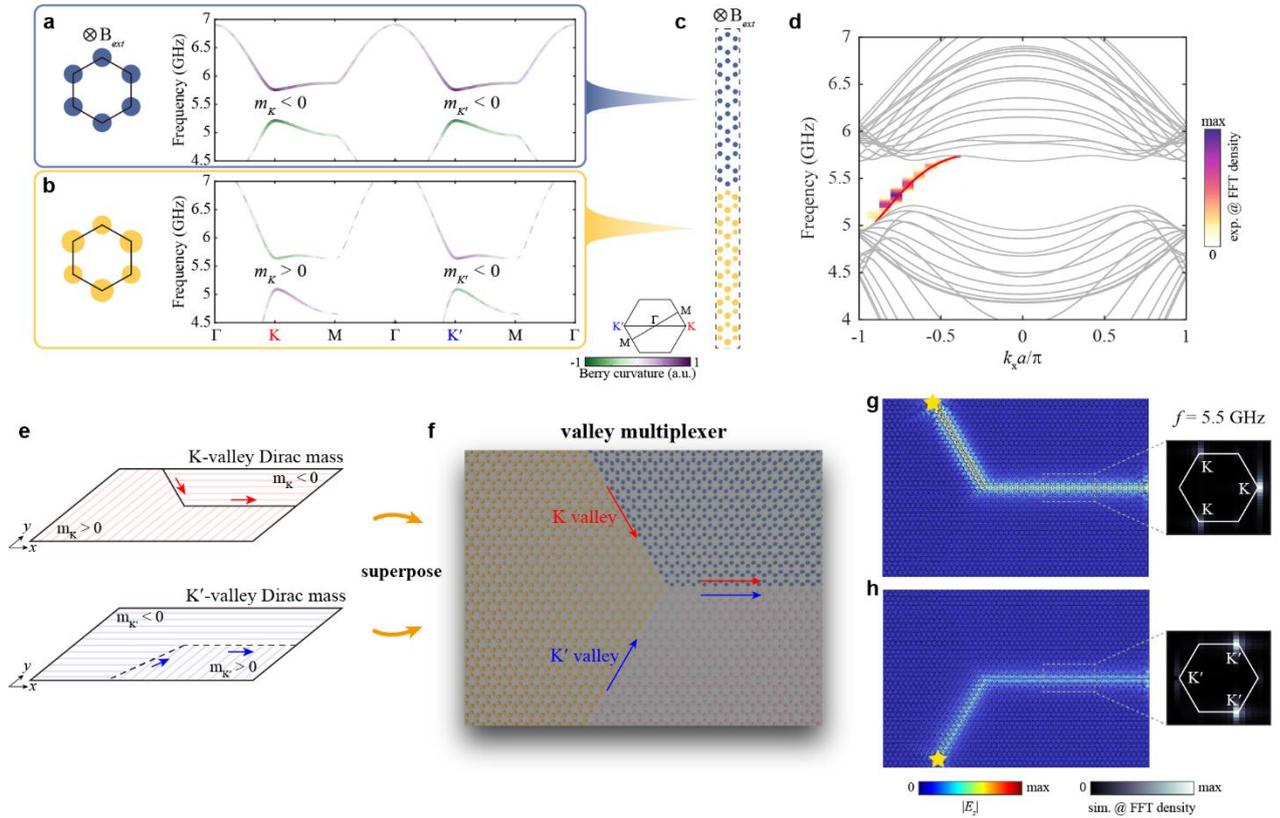

**Fig. 2 Photonic valley multiplexer. a** Bulk band structure of a Chern photonic crystal (CPC-1), with Berry curvature encoded as color, reflecting that CPC-1 has negative Dirac masses at both valleys. The unit cell of CPC-1 consists of two YIG rods under a biased external magnetic field. **b** Bulk band structure of a valley photonic crystal (VPC-1). VPC-1 has a positive Dirac mass at K valley and a negative Dirac mass at K' valley. Its unit cell is composed of two ceramic rods with different radii. **c** Photonic crystal waveguide supporting chiral K-valley edge states. **d** Edge dispersion of the chiral valley edge waveguide. The edge band is located around the K valley, confirming the exclusive support for K-polarized modes. Measured dispersion (FFT density) is represented by the background color. **e** Spatial distributions of $m_K(\mathbf{r})$ and $m_{K'}(\mathbf{r})$, forming two bending waveguides for K- and K'-valley waves, respectively. **f** Photonic valley multiplexer implemented as a Y-junction structure. This multiplexer is composed of three types of topological photonic crystals (CPC-1, CPC-2, and VPC-1), realized through the spatial superposition of mass distributions in (**e**). Notably, CPC-2 (the pink lattice) is the time-reversal counterpart of CPC-1 (with opposite external magnetic field), featuring positive Dirac masses at both valleys. **g,h** Simulated field profile when electromagnetic waves are incident from the upper or lower channel, respectively. Waves with different valley-polarizations converge into the horizontal bus waveguide without any valley flipping, verifying not only the functionality of valley multiplexer but also the one-way characteristic of valley chiral edge states.

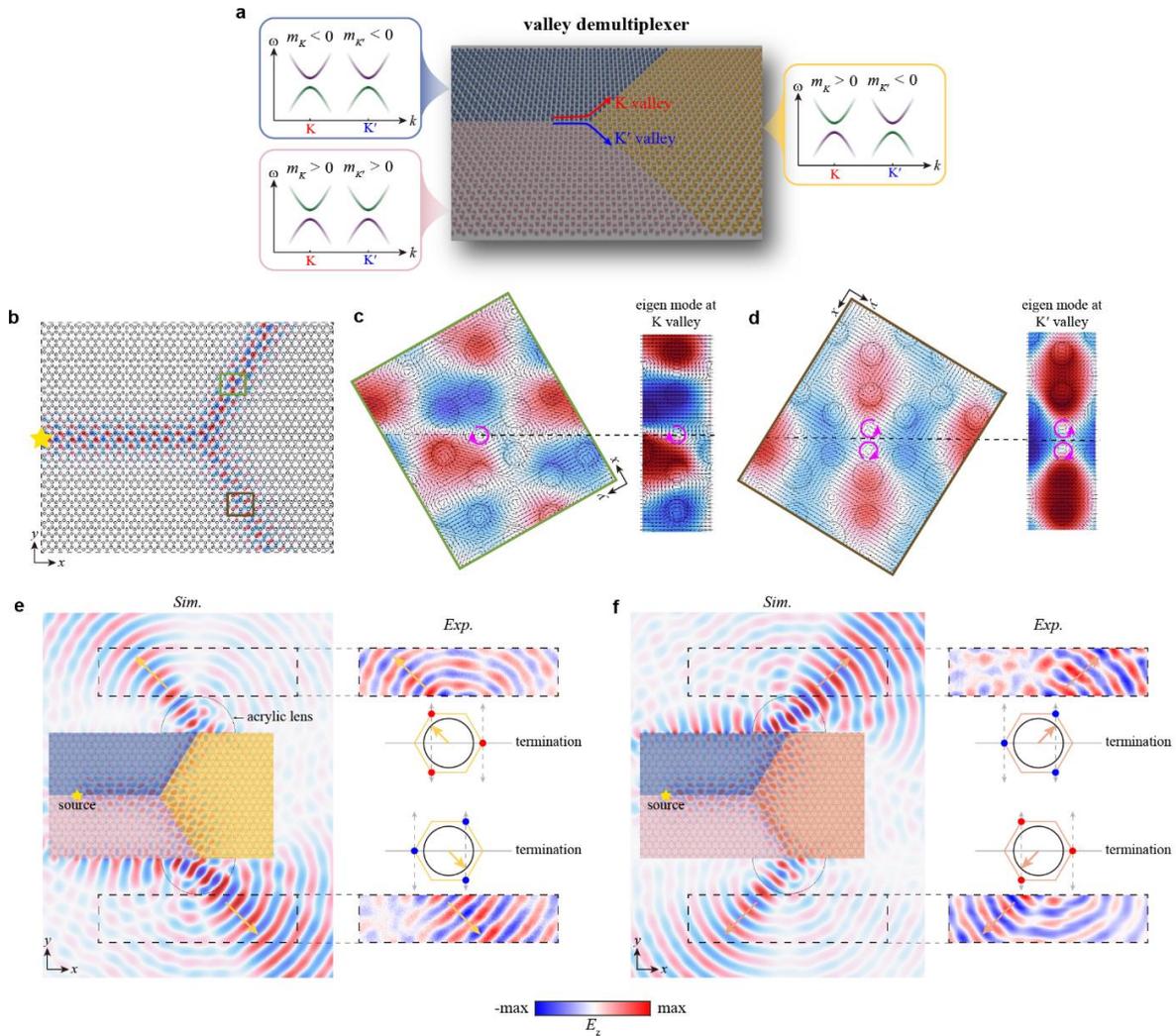

**Fig. 3 Photonic valley demultiplexer. a** Schematic diagram of a photonic valley demultiplexer. Insets: Valley Dirac masses of the three photonic crystals involved. **b** Simulated $E_z$-field profile at 5.5 GHz, excited by a point source located on the left (yellow star). **c** Left panel: $E_z$-field and Poynting vector distribution in the upper-right waveguide (framed in green box in (**b**)). Right panel: K-valley eigen edge mode between CPC-1 and VPC-1. The $E_z$-field and energy flow pattern of the propagating wave matches well with the eigen mode, confirming pure K-valley polarization. **d** Left panel: $E_z$-field and Poynting vector distribution in the lower-right waveguide (framed in brown box in (**b**)). Right panel: K'-valley eigen edge mode between CPC-2 and VPC-1. **e** Refraction of electromagnetic waves from the valley demultiplexer to air region at 5.57 GHz. Since the edge modes lie below the light cone of air, two acrylic semi-circular lenses (n = 1.45) are positioned at the two terminations to export the waves outward. Experimentally measured $E_z$ profiles are plotted in the right panel, aligning well with simulations. Inset: *k*-space analysis on the out-coupling waves, with black circles indicating the equifrequency contours of modes in the acrylic lenses. **f** Refraction of the electromagnetic waves in another valley demultiplexer where VPC-1 in (**e**) is replaced by VPC-2.

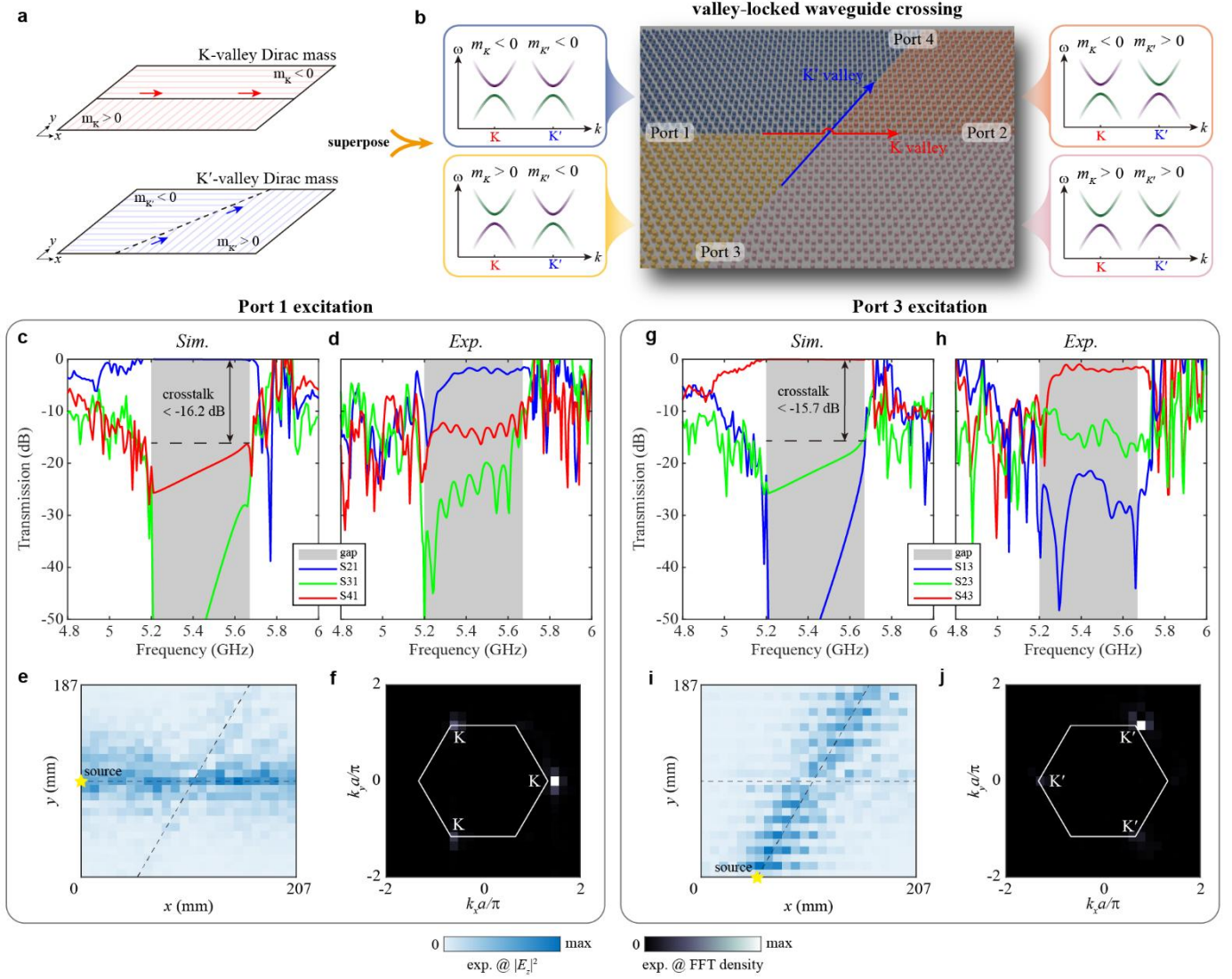

**Fig. 4 Valley-locked waveguide crossing. a** Spatial distributions of $m_K(\mathbf{r})$ and $m_{K'}(\mathbf{r})$, forming a horizontal channel for K-valley waves and a vertical channel for K'-valley waves. **b** Schematic of the valley-locked waveguide crossing, consisting of two CPCs (blue and pink) and two VPCs (yellow and orange). Insets: Valley Dirac masses of the four constituent photonic crystals. **c** Simulated transmission spectra of the waveguide crossing with electromagnetic waves excited at port 1. The crosstalk is characterized by the difference between S41 (red line) and S21 (blue line), signifying a crosstalk level below -16.2 dB. **d** Measured transmission spectra for port 1 excitation. **e** Measured $E_z$-field profile at 5.57 GHZ for port 1 excitation, demonstrating smooth propagation and negligible crosstalk. **f** Corresponding Fourier density of the field profile in (**e**), further confirming the valley-locked feature of the crossing. **g,h** Simulated and measured transmission spectra when electromagnetic waves are excited at port 3. **i,j** Measured $E_z$-field profile and corresponding Fourier density for port 3 excitation, confirming K'-valley polarization and low crosstalk.